\documentclass[11pt]{article}
\usepackage{amsmath,amssymb}

\usepackage{graphicx}
\usepackage{color}
\usepackage{caption}
\usepackage{float}
\usepackage{calligra}
\usepackage[
colorlinks=true,
linkcolor=blue,
urlcolor=blue,
filecolor=black,
citecolor=red,
]{hyperref}
\def\eg{{\it e.g. }}
\def\ie{{\it i.e.}}

\def\Tr{\mathrm{Tr}\,}
\def\p{\partial}
\def\bp{\bar{\partial}}

\def\half{\frac{1}{2}}
\newcommand{\be}{\begin{equation}}
\newcommand{\ee}{\end{equation}}

\newcommand{\bz}{\bar{z}}
\newcommand{\bx}{\bar{x}}
\newcommand{\by}{\bar{y}}

\newcommand{\CL}{{\mathcal{L}}}

\newcommand{\CO}{{\mathcal{O}}}
\newcommand{\CT}{{\mathcal{T}}}
\newcommand{\CK}{{\mathcal{K}}}
\newcommand{\CN}{{\mathcal{N}}}
\newcommand{\mt}{\tilde{\mu}}
\usepackage{calligra}

\DeclareMathAlphabet{\mathcalligra}{T1}{calligra}{m}{n}
\DeclareFontShape{T1}{calligra}{m}{n}{<->s*[2.2]callig15}{}
\newcommand{\scriptr}{\mathcalligra{r}\,}

\newcommand{\sr}{\scriptr}
\begin{document}
	\title{\textbf{Aspects of Holographic Entanglement Entropy for $T\bar{T}$-deformed CFTs}}

	\vspace{2cm}
	\author{ \textbf{Kuroush Allameh$^a$,  Amin Faraji Astaneh$^{a,b}$ and Alireza Hassanzadeh$^a$ }} 
	\date{}
	\maketitle
	\begin{center}
		\hspace{-0mm}
		\emph{$^a$ Department of Physics, Sharif University of Technology,\\
			P.O.Box 11155-9161, Tehran, Iran}
	\end{center}
	\begin{center}
		\hspace{-0mm}
		\emph{$^b$ School of Particles and Accelerators,\\ Institute for Research in Fundamental Sciences (IPM),}\\
		\emph{ P.O.Box 19395-5531, Tehran, Iran}
	\end{center}
	


	\begin{abstract}
		\noindent { We use the holographic methods to calculate the entanglement entropy for field theories modified by $T\bar{T}$ insertion. Based on the available holographic proposals, this calculation reduces to the holographic computations in AdS with finite cut-off. We perform a direct holographic calculation of the entanglement entropy for two dimensional deformed theory. By direct we mean the evaluation of the gravitational action in the bulk spacetime which has been reconstructed from a dual CFT$_2$\\ on $n$-sheeted Riemann surface as a finite cut-off boundary. We also apply Ryu-Takayanagi's prescription to investigate the most general structure of the entanglement entropy for $T\bar{T}$-deformed theories of arbitrary dimensions. In both cases, we find a perfect agreement with the known results.
		}
	\end{abstract}
	
	\vskip 8 cm
	\noindent
	\rule{7.7 cm}{.5 pt}\\
	\noindent 
	\noindent
	\noindent ~~~ {\footnotesize e-mails:\ k.allameh@physics.sharif.edu , faraji@sharif.ir , ar.hasanzadeh@physics.sharif.edu }


	\newpage
	\section{Introduction}
	Quantum Field Theories generally admit deformations that induce a flow at various energy levels. At high energy levels, the flow will be dominantly instigated by irrelevant operators. It would be very hard to probe the dynamics at such high energy levels mainly because that such deformed theories are generically non-integrable since we integrate out the high energy modes in the flow. Hence, it would be fair to ask if there is any integrable irrelevant deformation of field theories. Interestingly enough, Zamolodchikov has provided a positive answer: The $T\bar{T}$ deformation of a two-dimensional conformal field theory (CFT) is one of such deformations \cite{Zamolodchikov:2004ce}.

	$T\bar{T}$ operator is a composite irrelevant operator constructed from the holomorphic and anti-holomorphic parts of a two-dimensional CFT. Denoting the parameter of deformation by $\mu$, The $T\bar{T}$ deformed theory can be defined by the following flow equation
	\be
	\frac{\p\CL_\mu}{\p\mu}=\det(T_{ij})\, .
	\ee
	where $T_{ij}$ is the stress tensor of the deformed theory which in the complex coordinates decomposes to the holomorphic and anti-holomophic parts. The closed form of the Lagrangian can be iteratively fixed by solving the above equation. A unique feature of this irrelevant deformation is that not only it is integrable, but even more, it can improve the solvability of the original un-deformed theory \cite{Smirnov:2016lqw}, \cite{Cavaglia:2016oda}.
	
	A simple but rich enough example is the free Liouville theory on the complex plane with metric $ds^2=dxd\bx$,
	\be
	\CL_0=-\frac{c}{48\pi}\p\phi(x,\bx)\bp\phi(x,\bx)\equiv -\frac{c}{48\pi}\psi(x,\bx)\, ,
	\ee
	where $c$ stands for the central charge.\\
	The deformed Lagrangian takes the following closed form \cite{Bonelli:2018kik},\cite{Leoni:2020rof}\footnote{For us, the definition of the Liouville field differs by factor 2.} 
	\be\label{TbTCFT}
	\CL_\mu=\frac{2}{\mu}\left(\sqrt{1+\mu \CL_0}-1\right)\, .
	\ee
	The analysis may be continued to include the interaction terms as well.
	
	Interestingly, there is an intuitive holographic picture for $T\bar{T}$ deformation which is known as the \emph{finite cut-off proposal} \cite{McGough:2016lol}. Based on this holographic prescription, the deformed CFT is dual to a gravitational theory in AdS spacetime with finite cut-off. The latter condition is too crucial in this picture; the asymptotic boundary should be located at a finite radius instead of infinity. The finite radius will then match the parameter of deformation.\\
	There is another interesting holographic picture of $T\bar{T}$ deformation which comes from string theory on some UV/IR interpolating backgrounds \cite{Asrat:2017tzd}-\cite{Babaei-Aghbolagh:2020kjg}. This background has been verified to correspond to a two-dimensional CFT deformed by $T\bar{T}$.
	
To be more precise, let us focus on the energy spectrum of a $T\bar{T}$-deformed CFT on a cylinder of circumference $L$
		\be\label{ES}
		E_n=\frac{2\pi}{\mt L}\left(1-\sqrt{1-2\mt M_n+\mt^2 J_n^2}\right)\, ,
		\ee
		where $\mt=\frac{\pi\mu}{L^2}$ as defined in \cite{McGough:2016lol}.\\
		Here $M_n=\Delta_n+\bar{\Delta}_n-\frac{c}{12}$ and $J_n=\Delta_n-\bar{\Delta}_n$, where $(\Delta_n,\bar{\Delta}_n)$ are the conformal charges of the CFT state for which we compute the energy.
		
		Since we want to apply the holographic methods, we consider quantum field theories that admit holographic duals.  Referring to the AdS/CFT conjecture we know that a conformal field theory with gravity dual is well approximated by gravity at the large central charge limit \cite{Maldacena:1997re} and \cite{tHooft:1973alw}. But it is somehow a tricky limit for the deformed theories. To see this, let us simplify \eqref{ES} by setting $\Delta_n=\bar{\Delta}_n=0$ by which we arrive at the ground state energy. Then
		\be
		E_n=\frac{2\pi}{\mt L}\left(1-\sqrt{1+\frac{\mt c}{6}}\right)\, .
		\ee
		Obviously enough, while $c\rightarrow\infty$ is a completely safe limit when $\mt>0$, for negative sign, $\mt<0$ the large $c$ limit might be problematic when $c>\frac{6}{\mt}$. So it seems to be very important to individually deal with these two opposite signs.\footnote {It is worth noting that only the relative sign is meaningful since the individual signs differ here and there, due to the different conventions of various papers. Nevertheless, we follow the convention of \cite{McGough:2016lol}.}
		
		For the negative sign, it is possible to have a holographic picture considering the large $c$ limit and simultaneously tending $\mt$ to zero such that $\mt c$ remains fixed and smaller than 6.\\
		Various computations, such as the calculation of the entropy with the Cardy formula \cite{Cardy:1986ie}, suggest that the dual gravity is in fact a little string theory at the high energy limit which interpolates to the usual Einstein gravity in Anti-de-Sitter space at the low energy limits.
		
		But of course, the positive $\mt$ is also a legitimate signature that admits a clear holographic picture. While the large $c$ limit can be imposed without any problem for the positive sign of $\mt$, the only significant concern, in this case, is the contribution of the high energy modes to the energy spectrum. As it is evidenced by the explicit expression of the energy, it becomes imaginary for highly excited states. To avoid this, one needs to put a strong enough UV cut-off to block such high energy modes. In the dual picture, it means that the dual gravity should live in a segment of AdS which terminates at a finite cut-off behind which the high energy modes live.
	
	Besides the intuitive geometric picture of the deformation that such a  holographic recipe provides us, It enables us to generalize such irrelevant deformations to higher dimensions. It is natural then to try to construct a $T\bar{T}$-like composite operator from the Brown-York stress tensor in an asymptotically AdS$_{d+1}$ with a finite cut-off at the boundary \cite{Taylor:2018xcy} (see also \cite{Alishahiha:2019lng} as an example).
	
	This holographic construction significantly affects the pattern of the quantum correlation in quantum systems at the continuum limit. A very good measure to track this is the entanglement entropy (EE)\cite{Chakraborty:2018kpr}-\cite{Jeong:2019ylz} (See also \cite{Jafari:2019qns},  \cite{Chakraborty:2020fpt} for calculation of the quantum complexity). For this purpose, we consider the EE for two complementary sub-regions of a quantum mechanical system in the continuum limit. Naturally, The high energy modes are present in any arbitrary small region across the entangling surface. If this system admits a dual gravity then putting a UV finite cut-off in the geometry of bulk means introducing a lattice spacing scale above which we do not take the quantum correlation into account. The coincidence between the UV cut-off in AdS radial coordinate and the lattice minimum space in dual field theory is a direct consequence of the holography. We aim to explore how one could see the footprint of such a particular lattice scale on the pattern of the quantum entanglement of the system, specially when a finite cut off has been introduced in association with the $T\bar{T}$ deformation.
	
	Hence, our main goal in this note is to investigate some aspects of the entanglement entropy for deformed theories in two and higher dimensions using holography. We do this utilizing the direct holographic calculations, which is the subject of section \eqref{direct holography}.\\
	In section \eqref{minimal surface} we review the holographic calculation of the holographic entanglement entropy (HEE) for a two-dimensional CFT using the \emph{Ryu-Takayanagi}'s minimal surface prescription.
	Then we generalize the calculation to the most general cases of arbitrary dimensions and give a general form of the HEE for a generic deformed CFT. We examine some particular limits to show that our general results perfectly match with the known results in these special limits.
	
	The last section has been devoted to the conclusion.
	\section{Holographic Calculation of Renyi entropy for $T\bar{T}$-deformed CFT$_2$}\label{direct holography}
	For a composite quantum mechanical system that has been decomposed into two complementary parts, $A$ and $\bar{A}$, the EE is defined as
	\be
	S_{EE}=-\Tr\rho_A\log\rho_A\, ,
	\ee
	where $\rho_A=\Tr_{\bar{A}}\, \rho$ is the reduced density operator. This bi-partition in a field theory means the decomposition of the manifold on which the field theory lives, into parts with a common boundary $\Sigma$. this boundary will be known as the entangling surface. Instead of the direct calculation of the EE, it is more practical to define a Renyi entropy (RE) via replica method and extract the EE as a limit in replica number. More precisely, 
	\be
	S_n=\frac{1}{1-n}\log\Tr \rho_A^n \ , \ S_{EE}=\lim_{n\rightarrow 1}S_n\, .
	\ee
	Referring to \cite{Calabrese:2004eu} and \cite{Calabrese:2009qy}, in order to construct $\Tr\rho_A^n$ following the replica method one should first prepare $n$ copies of the manifold on which the original field theory lives. Gluing the copies along the cut at the location of the entangling surface in an appropriate way will generate the desired trace of the $n$'th power of the reduced density operator. As a result of this sewing of the manifolds, one gets a rather complicated geometry with conical singularities due to deficit angles in the direction of the Euclidean time. Such a geometry is called an $n$-sheeted Riemann surface, $\mathcal{R}_n$. 
	Therefore, the calculation boils down to the evaluation of the partition function on such a singular surface. Then
	\be
	S_{EE}=\lim_{n\rightarrow 1}\frac{1}{1-n}(\log Z_n-n\log Z_1)\ , \ Z_n=\Tr \rho_A^n\, .
	\ee
	
\begin{figure}[t]
		\begin{center}
			\includegraphics[width=0.6\linewidth]{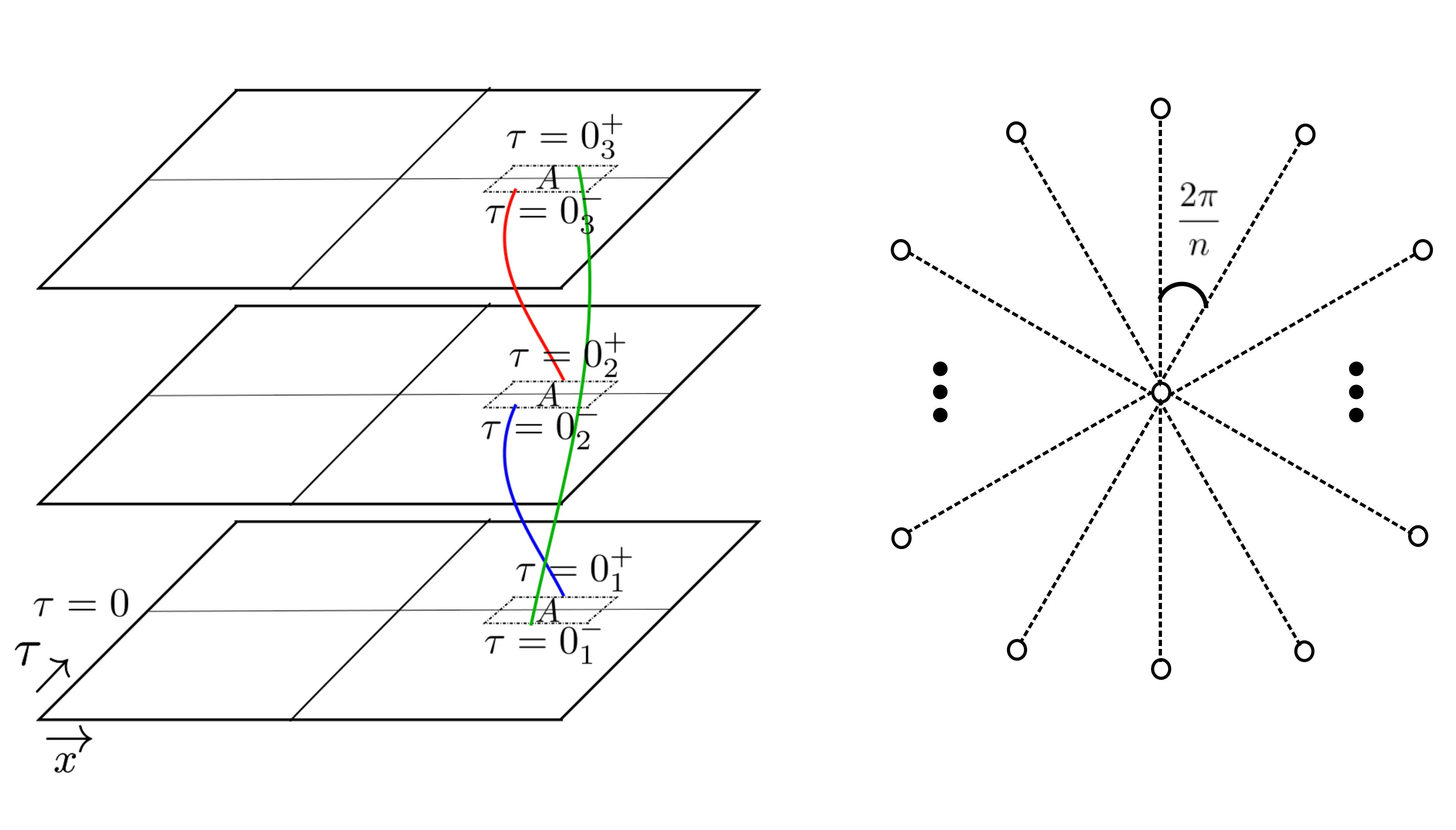}
		\end{center}
		\caption{$(n=3)$-sheeted Riemann surface (left). For illustrative purposes we chose the coordinates to be the Euclidean time $\tau$ and the spatial direction $x$. The colored lines represent the identifications along the cuts.
			In right hand side the complex plane with $(n+1)$ edge points has been depicted.}
		\label{fig}
	\end{figure}

	Referring to AdS/CFT correspondence one may substitute the logarithm of the partition function with the renormilzed gravitational action.
	In two dimensions we chose to work with complex coordinates $(x,\bx)$ to cover this Riemann surface on which the metric reads
	\be
	ds^2=dx d\bx\, .
	\ee
	As mentioned above, despite the simple form of the metric, this surface suffers from the conical singularities at the edges of the entangling surface, which is a spatial interval of length $\ell$, specified as $x\in[a,b]$.
	This non-triviality in geometry may be encoded in and extracted as a conformal factor if we move to the universal cover of $\mathcal{R}_n$ using the following transformation of the coordinates
	\be
	y=\left(\frac{x-a}{x-b}\right)^{1/n}\, .
	\ee
	Then 
	\be
	ds^2=\left\vert \frac{\p x}{\p y}\right\vert^2 dy d\by=e^{\phi(y,\by)}dyd\by\, ,
	\ee
	where
	\be
	e^{\frac{\phi(y,\by)}{2}}=n\ell\frac{\vert y\vert^{n-1}}{\vert y^n-1\vert^2}\, .
	\ee
	In this sense $\phi$ is the Liouville field.\\
	This transformation brings the $n$-sheeted Riemann surface to the complex plane on which the starting points of the interval go to zero and the end points will be distributed around a circle with infinitely large radius (see figure \eqref{fig}).
	
	Referring to the holographic reconstruction of spacetime \cite{deHaro:2000vlm} one may construct the bulk spacetime associated to discussed geometries on the boundary \cite{Hung:2011nu}.\\
	For this purpose, the proper form of the metric is the Fefferman-Graham metric \cite{FG}\footnote{We set the radius of AdS to 1 throughout this paper.}
	\be
	ds^2=\frac{d\rho^2}{4\rho^2}+\frac{1}{\rho}g_{ij}(\rho,X)dX^idX^j\ \ \ , \ \ \  g_{ij}(\rho,X)=g^{(0)}_{ij}(X)+\rho g^{(1)}_{ij}(X)+\cdots\, .
	\ee
	In our case $X^{1,2}=(y,\by)$ and $g^{(0)}_{ij}$ will be identified with $e^\phi dy d\by$. So the bulk metric can be reconstructed as follows \cite{Skenderis:1999nb}
	\be
	\begin{split}
		ds^2&=\frac{d\rho^2}{4\rho^2}+\frac{1}{\rho}e^\phi dy d\by+\frac{1}{2}\CT_\phi dy^2+\frac{1}{2}\bar{\CT}_\phi d\by^2+\frac{1}{4}R_\phi dy d\by\\
		&+\frac{1}{4}\rho e^{-\phi}(\CT_\phi dy+\frac{1}{4}R_\phi d\by)(\bar{\CT}_\phi d\by+\frac{1}{4}R_\phi dy)\, ,
	\end{split}
	\ee
	where
	\be
	\CT_\phi=\p_y^2\phi-\frac{1}{2}(\p_y\phi)^2\ \ \text{and} \ \ \bar{\CT}_\phi=\bp_y^2\phi-\frac{1}{2}(\bp_y\phi)^2\, ,
	\ee
	are the holomorphic and anti-holomorphic components of the Liouville field stress tensor respectively, and
	\be
	R_\phi=4\p_y\bp_y\phi\, ,
	\ee
	is related to the intrinsic curvature on the boundary.\footnote{In fact, $R_\phi=-e^\phi R^{(0)}$, where $R^{(0)}$ is the Ricci scalar for the metric $g^{(0)}_{ij}$.}
	
	Due to the symmetry group of AdS$_3$, there is a transformation to the Poincare coordinates \cite{Krasnov:2001cu}
	\be\label{FG to Poincare}
	r=\rho^{-1/2}e^{\phi/2}+\frac{1}{4}\rho^{1/2}e^{-\phi/2}\vert\p_y\phi\vert^2\ \ , \ \ z=y+\frac{1}{2}\, \frac{\rho e^{-\phi}\bp_y\phi}{1+\frac{1}{4}\rho e^{-\phi}\vert\p_y\phi\vert^2}\, .
	\ee
	which brings us to the following metric
	\be\label{PAdS}
	ds^2=\frac{dr^2}{r^2}+r^2 dzd\bz\, .
	\ee
	This transformation enables us to work with the usual Poincare metric in order to calculate the HEE. The only non-triviality is that the boundary is now a non-constant radius hypersurface with a non-trivial functionality of the coordinates.\\
	To see this let us focus on the near-boundary limit. In Fefferman-Graham case it is $\rho=\rho_c$. The radius of the finite cut-off, $\rho_c$ will be related to the parameter of the deformation, $\mu$. 
	Using the Brown-Hennaux relation, $c=\frac{3}{2G_N}$  and the convension used in \cite{McGough:2016lol}
	\be
	\mu=\frac{24\pi\rho_c}{c}\, .
	\ee
	Then if one keeps everything up to the order $\CO(\rho^{1/2})$, the coordinate transformation \eqref{FG to Poincare} yields $y=z$ and
	\be
	r_c=\frac{1}{\sqrt{\rho_c}}e^{\frac{\phi(z,\bz)}{2}}+\frac{1}{4}\sqrt{\rho_c} e^{-\frac{\phi(z,\bz)}{2}}\psi^2(z,\bz)\, .
	\ee
	So as we mentioned, the boundary in Poincare patch admits a coordinate dependent form specified as
	\be
	r_{bdy}(r,z,\bz)=r-r_c(z,\bz)=0\, .
	\ee
	Now we can start to evaluate the renormalized gravitational action
	\be
	\begin{split}
		I_r&=\frac{1}{16\pi G_N}(I_{EH}+I_{GH}+I_{ct})\, ,\\
		&I_{EH}=\int d^3X\sqrt{-g}(R+2)\, ,\\
		&I_{GH}=2\int d^2X\sqrt{-h}K\, ,\\
		&I_{ct}=-2\int d^2X\sqrt{-h}(1+\rho_c\, f(z,\bz))\, .
	\end{split}
	\ee
	
	It is worth noting that the deformed theory needs a subtle regularization since the usual counterterm of the gravitational action has been designed to remove the divergent terms and thus there always remains some sorts of ambiguities due to the possible addition of the finite terms. But now these finite terms take a meaningful role as a natural consequence of the movement of the boundary to a finite radius. One observes that making a reasonable connection to the $T\bar{T}$ deformation, the legitimate finite terms of order $\rho_c$ which can contribute to the on-shell action may only come from the bulk. So we chose $f(z,\bz)$ such that no boundary terms of order $\rho_c$ remain in the boundary action. The ambiguity in the counterterm has been pointed out also in \cite{Chen:2018eqk}, however, to the best of our knowledge, the systematic way of distinguishing the proper regularization scheme has not been explored in the literature as of yet. We call it the \emph{Maximal Boundary Subtraction} (MBS) scheme of regularization and we believe that the proposed scheme of regularization can be vastly applied to the variations of the two dimensional problem.
	
	In the case of AdS$_3$, $R=-6$ and the EH action gives
	\be
	I_{EH}=-\frac{1}{\rho_c}\int e^\phi\, dzd\bz -\frac{1}{2}\int\psi\, dzd\bz-\frac{\rho_c}{16}\int e^{-\phi}\psi^2\, dz d\bz
	\ee
	While on the boundary
	\be
	I_{GH}+I_{ct}=\frac{1}{\rho_c}\int e^\phi\, dzd\bz-2\int \p\bp\phi\, dzd\bz\, ,
	\ee
	Interestingly enough, $\phi(z,\bz)$ will be decomposed in holomorphic and anti-holomorphic parts
	\be
	\phi(z,\bz)=\log Z(z)\bar{Z}(\bz)+c\, ,
	\ee
	which guarantees that 
	\be\label{eqm1}
	\p_z\bp_z\phi=0\, .
	\ee
		Reexpressing the prefactors in terms of the central charge and the parameter of deformation we finally get \footnote{The authors of \cite{Chen:2018eqk} also obtain the renormalized action to show that the gravitational on-shell action gives the partition function of the deformed field theory at the linear order of perturbation. It is to take care of the back reaction of the twist operators as they have mentioned in their paper. In field theory side they calculate the replicated partition function taking the $T\bar{T}$ as an irrelevant insertion on the $n$-sheeted Riemann surface.}
	\be
	I_r=-\frac{c}{48\pi}\int dzd\bz\left[ \psi(z,\bz)+\frac{\mu c}{192\pi}e^{-\phi(z,\bz)}\psi^2(z,\bz)\right]\, .
	\ee
	using \eqref{eqm1} one may easily write
	\be
	e^{-\phi}\psi^2=(\p\bp e^{-\phi/2})^2\, .
	\ee
	It would be possible then to rewrite the renormalized action in the following covariant form 
	\be
	I_r=-\frac{c}{96\pi}\int dV_2 \left[\p_i\phi\p^i\phi+\frac{\mu c}{48\pi}(\p_i\p^i e^{-\frac{\phi}{2}})^2\right]\, .
	\ee
Integrating by part one may reduce the integral to a boundary term with the use of the equations of motion. By boundary term, we mean the integral over the boundaries of the $n+1$ punctures on the complex plane. The punctures which from now on will be denoted by $P_a, a=1,2,\cdots n+1$, are small circles that determine short distance cut-off across the entangling surface, as usual, see figure \eqref{fig}. We naturally identify this cut-off with the finite radial cut-off and thus up to a factor with the parameter of deformation. As mentioned above, the renormalized action turns out to be a boundary integral over the boundaries of the punctures, $\p P_a$
	\be
	I_r=-\frac{c}{96\pi}\int_{\p P_a} dS_n\left[\phi\p_n\phi+\frac{\mu c}{48\pi}\left(\p_ne^{-\frac{\phi}{2}}\Box e^{-\frac{\phi}{2}}-e^{-\frac{\phi}{2}}\p_n\Box e^{-\frac{\phi}{2}}\right)\right]\, ,
	\ee
	where $dS_n$ is the integrating measure on the boundary and $\partial_n=n^i\partial_i$ is the normal derivative. The first term of the integral gives \cite{Hung:2011nu}  (see also \cite{Lunin:2000yv}, \cite{FarajiAstaneh:2013oey}),
	\be
	I_r^{(n)}=\frac{c}{6}\left(\frac{n^2-1}{n}\right)\log\left(\sqrt{\frac{24\pi}{\mu c}}\ell\right)\, .
	\ee
	 To evaluate this integral we adopt the polar coordinates $z=\vert z\vert e^{i\theta}$. Then
	\be
	e^{-\frac{\phi}{2}}=\frac{1}{n\ell}(\vert z\vert^{n+1}+\vert z\vert^{-n+1}-2\vert z\vert\cos n\theta),
	\ee
	and one finds
	\be
	\int_{\p P_a} dS_n\, \left(\p_ne^{-\frac{\phi}{2}}\Box e^{-\frac{\phi}{2}}-e^{-\frac{\phi}{2}}\p_n\Box e^{-\frac{\phi}{2}}\right)=-32\pi(\frac{n^2-1}{n^2})\frac{1}{\ell^2}\, .
	\ee
	Therefore finally we get
	\be
	I_r^{(n)}=\frac{c}{6}\left(\frac{n^2-1}{n}\right)\left[\log\left(\sqrt{\frac{24\pi}{\mu c}}\ell\right)+\frac{\mu c}{24\pi n\ell^2}\right]\, .
	\ee
	and thus
	\be\label{EE}
	S_{EE}=\frac{c}{3}\log\left(\sqrt{\frac{24\pi}{\mu c}}\ell\right)+\frac{\mu c^2}{72\pi\ell^2}\, .
	\ee
	This result perfectly matches with the result of the calculation on CFT side, see for instance equation (2.9) in \cite{Donnelly:2018bef}.\footnote{Note that the radius $r$ in \cite{Donnelly:2018bef} is equal to $\ell/2$ in our work.}
	
	It is worth making a comment on the entropic c-function here. The Casini-Huerta entropic \\c-function in two dimensions is defined as \cite{Casini:2012ei}
		\be
		C(\ell)=\ell \frac{dS_{EE}}{d\ell}\, .
		\ee
		They propose some sort of entropic c-theorem by which they in fact rephrase the Strong Sub-Additivity inequality (SSA). Based on this theorem, the length of the entangling interval stands for the RG scale and $\frac{dC(\ell)}{d\ell}\leq 0$.
		For a $T\bar{T}$ deformed CFT, the entropic c-function depends on the parameter of deformation and thus equivalently to the radius of the finite cut-off
		\be
		C(\ell)=\frac{c}{3}-\frac{\mu c^2}{36\pi\ell^2}\, .
		\ee
		So for a $T\bar{T}$-deformed CFT one has $\frac{dC(\ell)}{d\ell}> 0\, .$
		Such violence has roots in the volume law of the deformed part of the EE. It is known that SSA is substantially sensitive to the locality of the theory across the entangling sub-region. A strong sign of locality is the area law of the EE which is faded out for the $T\bar{T}$ deformed part \cite{Lewkowycz:2019xse}.
		Similar behavior will be observed for the deformation with the opposite sign of $\mu$, \cite{Chakraborty:2018kpr} and \cite{Asrat:2019end}.
		 
		One should also note that the gravitational on-shell action with finite cut-off and evaluated for the replicated geometry provides us more information besides the EE. For example, one may compute the ``Modular Entropy'' which interestingly admits a holographic area law for the cosmic branes \cite{Dong:2016fnf}. \\
		One can also start with our finding and derive the higher cumulants of the entropy rather than the EE. An interesting example in which our direct holographic calculation might be useful in the calculation of ``Capacity of Entanglement'' which can tell us about the quantum gravitational fluctuations around the Ryu-Takayanagi surface \cite{DeBoer:2018kvc}.
	\section{Holographic Ryu-Takayanagi's minimal surface prescription}\label{minimal surface}
	In this section, we apply Ryu-Takayanagi's minimal surface prescription to calculate the HEE in AdS spacetime with a finite cut-off.\\
	Based on the original version of this proposal \cite{Ryu:2006bv}, \cite{Ryu:2006ef}
	\be
	S_{EE}=\mathrm{Min}\frac{A_\gamma}{4G_N}\, ,
	\ee
	where $A_\gamma$ is the area of a holographic codimension-two surface homologous to the entangling surface $\Sigma$. As mentioned, the minimality condition should be imposed to get the desired result.
	Applying this proposal to the deformed theory in two dimensions, one needs to evaluate the length of a geodesic denoted by $\gamma$ in a time slice of \eqref{PAdS}, which at the finite cut-off boundary meets the entangling interval. In this case, the turning point at which the radial derivative changes sign will be obtained as a function of the finite cut-off of radius $r_c$. We have
	\be
	r_t=\frac{r_c}{\sqrt{1+\frac{\ell^2 r_c^2}{4}}}\, .
	\ee
	Then the length of the geodesic reads (See also \cite{Donnelly:2018bef})
	\be
	A_\gamma=2\log\left(\frac{\ell r_c}{2}+\sqrt{1+\frac{\ell^2 r_c^2}{4}}\right)=2\log\left(\ell r_c\right)+\frac{2}{\ell^2 r_c^2}+\cdots\, .
	\ee
	So one can easily derive the HEE which perfectly matches \eqref{EE}.
	One can generalize this calculation to higher dimensions. To do so we consider a spherical entangling surface of codimension-two on the (finite cut-off) boundary. Writing down the AdS metric
	\be
	ds^2=\frac{dr^2}{r^2}+r^2(d\tau^2+d\sr^2+\sr^2d\Omega_{d-2}^2)\, .
	\ee
	The RT surface, $\gamma$ will be parameterized as $\tau=0\, , \, \sr=\sr(r)\, .$ Solving the minimality condition one gets
	\be
	\sr(r)=\sqrt{R^2-\frac{1}{r^2}}\, .
	\ee
	Then the area of the minimal surface reads
	\be
	A_\gamma^{min}=\Omega_{d-2}\int^{r_c} dr\sqrt{1+r^4\sr'(r)^2}r^{d-3}\sr(r)^{d-2}\, .
	\ee
	Focusing on the cut off dependent terms, the HEE reads (See \cite{Park:2018snf}, \cite{Banerjee:2019ewu} fo similiar discussion)
	\be\label{EE for S}
	S_{EE}=\frac{\Omega_{d-2}}{4G_N}\left(\frac{R^{d-2}}{d-2}r_c^{d-2}-\frac{R^{d-4}}{2(d-4)}r_c^{d-4}+\cdots\right)
	\ee
	One should note that we have chosen to write the expression in terms of the radius of the entangling surface on the finite cut-off surface, $r=r_c$. Denoting this radius again by $R$, we have imposed the replacement $R\rightarrow \sqrt{R^2+\frac{1}{r_c}}$ in the final step.\\
	For $d=2$, substituting the radius $R$ with $\ell/2$ one reproduces the excess of the EE due to the $T\bar{T}$ deformation, \ie\,  the second term of \eqref{EE}.
	Another simple interesting example is the EE in $d=3$ for which
	\be
	S_{EE}=\frac{\pi}{2G_N}(Rr_c+\frac{1}{2Rr_c})\, .
	\ee
	The first term is an area term and the second term captures the information of the $T\bar{T}$-like deformation in three dimensions. The latter term depends on the inverse of the radius of the entangling surface. Here we have concentrated on a cut-off dependent term. There is a constant cut-off independent term originating from the free energy of CFT on the Euclidean sphere $S^3$ which should be added to the entropy as well \cite{Casini:2012ei}. Obviously, in four dimensions the second term in \eqref{EE for S} becomes a logarithm.
	
	We can perform even more general calculations and construct the leading terms of the EE in a finite cut-off AdS$_{d+1}$ for the most general cases. To do so we use the Fefferman-Graham metric again, but now in $d+1$ dimensions. See \cite{Graham:1999pm}-\cite{Astaneh:2014uba} for the calculation of the area of a submanifold in AdS spacetime.
	
	We aim to calculate the holographic entanglement entropy for a generic entangling surface. To do so we chose the metric on the boundary to be of the form of $S_2\times\Sigma_{d-2}$, \cite{Fursaev:1995ef}-\cite{FarajiAstaneh:2014oju}.
	\begin{equation}
	\begin{split}
	ds^2&=g^{(0)}_{AB}dX^AdX^B=(1-\frac{1}{4}R^{(0)}_{abab}\sr^2+\cdots)(d\sr^2+\sr^2d\tau^2)\\
	&+\left\{h_{ij}(x)+2\sr K^a_{ij}(x)n^a+\half\sr^2[ (K^aK^a)_{ij}- R^{(0)}_{aiaj}]+\cdots\right\}dx^idx^j\, .
	\end{split}
	\end{equation}
	Here $\tau$ is the Euclidean time and $\sr=0$ defines the locations of the entangling surface $\Sigma$ which is covered by coordinates $x^i$. \\
	$\Sigma$ is codimension-two and therefore there are two unit normal vectors on that denoted by $n^a$ where $a=1,2$ correspond to $\tau$ and $r$ directions, respectively.
	In our notation $R^{(0)}_{abab}$ denotes the curvature tensor contracted with the unit normal vectors and we use the summation conventions for the normal vectors, \eg $K^a_{ij}(x)n^a=\sum_{a=1,2}K^a_{ij}(x)n^a$.\\
	We want to evaluate the area of a minimal surface in the bulk denoted by $\gamma$. This surface will be parameterized by
	\be
	\sr=\sr(\rho)=\sum_{k=1,2,\cdots}\sr_k(\rho-\rho_c)^k\, ,
	\ee
	where again, $\rho_c$ is a small but finite cut-off. The coefficients $\sr_k$ will be determined by imposing the minimality condition, $\Tr\CK_\gamma^a=0$. To get the leading and subleading terms, it is sufficient to terminate the expansions at the order $\CO(\sr^2)$ or equivalently $\CO(\rho)$. The area element of the holographic hypersurface $\gamma$, reads then
	\be
	\begin{split}
		d A_\gamma&=\frac{1}{2\rho^{d/2}}\sqrt{1+4\rho \sr'(\rho)^2}(1-\sr(\rho) \Tr K+\frac{1}{2}\rho\Tr g^{(1)})d A_\Sigma\,  d\rho\, ,
	\end{split}
	\ee
	where we have set $K^1=0$  due to the symmetry in the time direction and we have chosen the unit normal in $r$ direction to be outward. We should emphasize that the traces are defined in terms of the induced metric on $\Sigma$ in the above expression. Thus
	\be
	\Tr g^{(1)}=\frac{1}{d-2}\left(R_{aa}-\frac{d}{2(d-1)}R\right)\, .
	\ee
	To determine the unknown coefficient $r_1$ we use the minimality condition, $\Tr\CK^{1,2}_\gamma=0$.\\
	$\Tr\CK^{1}_\gamma=0$, because of the symmetry in the time direction, but the second condition is nontrivial to solve. To find it we first note that there are two normal vectors on $\gamma$ in spatial directions
	\be
	\CN_\sr=\frac{1}{\sqrt{g^{\sr\sr}+g^{\rho\rho}\sr'^2}} \ \ , \ \ \CN_\rho=\frac{-\sr'}{\sqrt{g^{\sr\sr}+g^{\rho\rho}\sr'^2}}\ ,
	\ee
	where $g^{\rho\rho}=4\rho^2$ and $g^{\sr\sr}=\rho(1+\frac{1}{4}\sr^2R_{abab})+\cdots$.\\
	Then we will have
	\be
	\Tr \CK^{(2)}_{\gamma}=\frac{1}{\sqrt{G}}\p_\sr(\sqrt{G}\CN^\sr)+\frac{1}{\sqrt{G}}\p_\rho(\sqrt{G}\CN^\rho)=0\, ,
	\ee
	Which can be solved to find
	\be
	\sr_1=\frac{\Tr K}{2(d-2)}\, .
	\ee
	So finally up to the subleading terms we get
	\begin{equation}
	A_{\gamma}^{min}=\frac{A_\Sigma}{(d-2)\delta^{d-2}}+\frac{1}{2(d-2)(d-4)\delta^{d-4}}\int_\Sigma dv_\Sigma \left[R_{aa}-\frac{d}{2(d-1)}R-\frac{1}{d-2}(\Tr K)^2\right]\ .
	\end{equation}
	Although we have calculated the entropy for the general case in a finite cut-off AdS, in order to make the connection to deformed CFT we restrict ourselves to the flat base manifold with the symmetry in time direction. Then
	\be
	S_{EE}=\frac{r_c^{d-2}}{4(d-2)G_N}A_\Sigma-\frac{r_c^{d-4}}{2(d-2)^2(d-4)G_N}W_\Sigma\, ,
	\ee
	where
	\be
	W_\Sigma=\frac{1}{4}\int_\Sigma (\Tr K)^2\, ,
	\ee
	is the Wilmore energy of the entangling surface \cite{Willmore}.\footnote{Here $\Tr K$ stands for the trace of extrinsic curvature tensor in spatial directions.}\\
	This is our final result for the leading terms of the entanglement entropy with finite cut-off. 
	For a spherical entangling region of radius $R$,
	\be
	W_\Sigma=\frac{(d-2)^2\Omega_{d-2}}{4}R^{d-4},
	\ee
	and one rearrive at \eqref{EE for S}. It is worth noting that in the usual cases when we calculate the HEE for AdS with boundary at infinity, the Willmore term comes with a factor $(d-3)$. In AdS with finite cut-off this factor drops out and we find a contribution to the EE of order $r_c^2$ (or $\mu$ in the notation of deformed field theory) in three dimensions. In four dimensions the correction would be logarithmic in terms of the parameter of the deformation.
	We will discuss some more details in conclusion.
	
	\section{Conclusion}
	In this paper, we have investigated some aspects of the HEE for a $T\bar{T}$-deformed two-dimensional CFT and the similar theories of higher dimensions. In two dimensions, the deformation is well defined and the closed forms of the Lagrangian with interactions are well studied in the literature. We have used the ``AdS with finite cut-off'' proposal to calculate the EE for such deformed theories, holographically. It became possible by choosing a non-constant radius finite cut-off in Poincare coordinates as mentioned in section \ref{direct holography}. The profile of this surface will be fixed by the replica construction of the boundary theory. We have calculated the gravitational on-shell action in this setup and applied the AdS/CFT correspondence directly to derive the Renyi entropy and then the EE of deformed theory.
	
	One of the most important parts of this calculation was the regularization procedure of the action. Since the role of the counterterms is to remove the divergences in a covariant and appropriate way, they are always ambiguous, because of the possible finite terms which can be added to the boundary action. One usually chooses the minimal subtraction scheme of regularization to remove just the divergent terms using the counterterms. Now, in the case of the $T\bar{T}$ deformation and its holographic dual, such finite terms find indispensable meaning, and thus it is very important to be careful about what counterterms we should take into account to remove the divergences without touching the content of the deformed theory. We observed that the finite terms which correspond to the $T\bar{T}$ deformation come from the bulk term and thus we arrange the counterterms in a way that neither divergent terms nor finite boundary terms at the order at which deformation shows up, contribute to the on-shell action. We called it the Maximal Boundary Subtraction (MBS) scheme of regularization. 
	
	Besides this direct holographic calculation, we have reviewed the derivation of the EE for the deformed theories using Ryu-Takayanagi's minimal surface prescription. We find a perfect match with the previous holographic setup and of course with the already known results in field theory.
	\\
	We have generalized this holographic calculation to higher dimensions by calculating the leading correction to the entropy due to the finite cut-off. While the leading term exhibits an area law, the correction is proportional to the Willmore energy of the entangling surface for deformed CFT. According to the discussion in \cite{Astaneh:2014uba} there are global extremizers for this function in any dimension and for various classes of topology. Spherical regions are supposed to be the extremizers for the class of surfaces of zero genera (see \cite{Willmore}, \cite{W2} and \cite{Astaneh:2014uba}). So in this category, the spheres capture the maximal information of the $T\bar{T}$-like deformations among the surfaces with the same areas. It would be interesting to investigate this aspect in more details.
	
	A very interesting question about the EE is that how the excitation modifies the pattern of the quantum correlation in a given CFT. This has been investigated for usual CFTs but not very extensively for $T\bar{T}$ deformed CFTs. Our direct holographic calculation is very instructive for this purpose. Using the AdS/CFT correspondence one needs just to add matter fields to the bulk action which stand as sources for insertions on the boundary deformed theory. The subtle regularization method which we have introduced in our manuscript is very crucial for such investigation. This is a project in progress from which we think a thermodynamics for $T\bar{T}$ deformation would finally emerge.
	
	In this paper, we have concentrated on the Einstein theory of gravity. It would be interesting to consider gravitational theories with higher derivative terms in finite cut-off AdS spacetime. In particular it is known that massive theories of gravity admit logarithmic solutions at some special limits. It is worth investigating the $T\bar{T}$ deformation of such logarithmic field theories.
	
	The deformation of the boundary conformal field theories (BCFT) is also of great interest. There are interesting holographic pictures of BCFTs \cite{Takayanagi:2011zk}-\cite{FarajiAstaneh:2017hqv}, which incorporating them with finite cut-off proposals enables us to understand how the bulk stress tensors and the boundary stress tensors come into play in the case of the $T\bar{T}$ deformation of a theory with boundary. These problems are currently under investigation.
	
	\section*{Acknowledgement}
	We would like to thank D. Kutasov, A. Naseh and S. Solodukhin for useful comments.

\end{document}